# Measurement of parity-dependent energy-phase relation of the low-energy states in a potential artificial Kitaev chain utilizing a transmon qubit

Enna Zhuo[1,2,†], Xiaozhou Yang[1,2,†], Yuyang Huang[1,2,†], Zhaozheng Lyu[1,2,5,†,*], Ang Li[3], Bing Li[1,2], Yunxiao Zhang[1,2], Xiang Wang[1,2], Duolin Wang[1,2], Yukun Shi[1,2], Anqi Wang[1,2], E.P.A.M. Bakkers[4], Xiaodong Han[3], Xiaohui Song[1,2,6], Peiling Li[1], Bingbing Tong[1], Ziwei Dou[1], Guangtong Liu[1,2,5,6], Fanming Qu[1,2,5,6], Jie Shen[1,2,6], and Li Lu[1,2,5,6,*]

[1] *Beijing National Laboratory for Condensed Matter Physics, Institute of Physics, Chinese Academy of Sciences, Beijing 100190, China*

[2] *School of Physical Sciences, University of Chinese Academy of Sciences, Beijing 100049, China*

[3] *Beijing Key Laboratory of Microstructure and Property of Advanced Materials, Beijing University of Technology, Beijing 100124, China*

[4] *Department of Applied Physics, Eindhoven University of Technology, 5600 MB Eindhoven, The Netherlands*

[5] *Hefei National Laboratory, Hefei 230088, China*

[6] *Songshan Lake Materials Laboratory, Dongguan, Guangdong 523808, China*

[†] These authors contribute equally to this work.

[*] Corresponding authors: Zhaozheng Lyu: lyuzhzh@iphy.ac.cn, Li Lu: lilu@iphy.ac.cn

## Abstract

Artificial Kitaev chains have emerged as a promising platform for realizing topological quantum computing. Once the chains are formed and the Majorana zero modes are braided/fused, reading out the parity of the chains is essential for further verifying the non-Abelian property of the Majorana zero modes. Here we demonstrate the feasibility of using a superconducting transmon qubit, which incorporates an end of a four-site quantum dot-superconductor chain based on a Ge/Si nanowire, to directly detect the singlet/doublet state, and thus the parity of the entire chain. We also demonstrate that for multiple-dot chains there are two types of 0-$\pi$ transitions between different charging states: the parity-flip 0-$\pi$ transition and the parity-preserved 0-$\pi$ transition. Furthermore, we show that the inter-dot coupling, hence the strengths of cross Andreev reflection and elastic cotunneling of electrons, can be adjusted by local electrostatic gating in chains fabricated on Ge/Si core-shell nanowires. Our exploration would be helpful for the ultimate realization of topological quantum computing based on artificial Kitaev chains.

# Introduction—

Majorana zero modes in solid-state systems have received considerable attention, since they hold the promise of realizing topological quantum computing which is immune to environmental noise [1-5]. To host Majorana zero modes, various platforms have been explored, such as those based on intrinsic *p*-wave superconductor candidates [6-8] and/or superconductor-semiconductor heterostructures with *p*-wave-like superconductivity [9, 10]. Recently, the creation of an artificial Kitaev chain (AKC) utilizing a series of superconductor-mediated spin-polarized quantum dots (QDs) has emerged as a highly sought-after objective in Majorana physics researches [11-13], thanks to its potential to withstand material disorder [14, 15]. Two-QD [16-18] and three-QD [19] AKCs that support Majorana zero modes have been successfully implemented. These AKCs are anticipated to be the building blocks for Majorana-based qubits, with the parity information encoded in the Majorana pairs at the ends of the chains [20-22]. Along this direction, one of the next experimental challenges is how to read out the parity of the chains. The parity information may be reflected on the charging energy [23] and/or the Josephson energy [24] of the systems. Fast parity readout may be realized in tunneling measurements [25, 26], quantum capacitance measurements [27-29], and circuit quantum electrodynamics (cQED) measurements [30-36]. Among them, cQED stands out for its superior energy and time resolutions. Also, no quasiparticles will be injected to the systems during the measurement.

In this study, we constructed a four-site superconductor-quantum dot (SC-QD) chain based on a Ge/Si core/shell nanowire, and created a superconducting transmon by incorporating an end of the chain. This hybrid device enabled us to determine the lowest-energy states and the parities of the chain using a cQED technique. Utilizing this technique, we found both parity-flip and parity-preserving 0-π transitions in the chain. And, we demonstrated the gate-voltage control of inter-dot couplings, an ability which is of crucial importance for the creation of Majorana zero modes in the chains.

# The four-site SC-QD chain—

We fabricated a four-site $SC_1$-$QD_1$-$SC_2$-$QD_2$ chain by depositing Al electrodes onto a Ge/Si core/shell nanowire. The quantum dots were adjusted to desired sizes by diffusing Al into the nanowire in a controlled annealing process during which the resistance of the nanowire was continuously monitored [37]. Such a device, depicted in Fig. 1(a), facilitates the coupling of Andreev bound states between two separate dots through both cross Andreev reflection (CAR) and elastic cotunneling (ECT) of electrons. In the CAR process, an incoming electron is reflected from one dot as a hole into the other dot, as indicated by the purple arrows in the top panel of Fig. 1(a). Conversely, the ECT involves the transfer of a single electron from one dot to another, as indicated by the red arrow in the top panel of Fig. 1(a). The relative strength of CAR and ECT determines the detailed shape of the charge stability diagram, and the balance between them determines the appearance of Majorana zero modes.

Figure 1(b) shows the charge stability diagram of the chain. The charge depletion state (0,0) is determined experimentally, and the rest part of the diagram is deduced by modeling the chain as four single-spinful-orbital sites with nearest-neighbor hopping $t_i$ ($i = 1, 2, 3$) and next-nearest-neighbor hopping $\tau$, both depicted in the bottom panel of Fig. 1(a) (the details of the model is described in the Supplemental Material). The diagram is divided into distinct regions, each labeled with the occupation numbers of the electrons in QD$_1$ and QD$_2$, $(n_{\text{QD}_1}, n_{\text{QD}_2})$, where $n_{\text{QD}_1}, n_{\text{QD}_2} \in \{0, 1, 2\}$. The gray line denotes the parity boundary when $\phi = \pi$, which is slightly different from that in the $\phi = 0$ case. In our device, the two QDs are controlled by a common bottom gate, which allows tuning along only one electrostatic direction. The gate-tuning path indicated by the brown line in Fig. 1(b) is inferred by matching the evolution of the measured transmon spectra with numerical simulations, and is consistent with all key spectral signatures observed along the tuning range. Although the exact path cannot be uniquely determined due to charge trapping and hysteresis, the overall trajectory enables qualitatively reliable assignments of charge occupations. Details of the path determination are provided in the Supplemental Material.

Figure 1(c) further depicts the expected energy-phase relations (EPRs) of the chain for the four different charging states in the charge stability diagram. For this two-QD chain, when the total fermion parity of the chain is even (odd), the ground state is expected to be a singlet (doublet). And, when the occupancy number of the measured QD (QD$_1$ in this case) is even (odd), the expected minimum energy of the EPR is located at or near $\phi = 0$ ($\pi$). Specially, the EPR in the (1,1) region contains a spin-singlet state $(|\uparrow,\downarrow\rangle - |\downarrow,\uparrow\rangle)/\sqrt{2}$ and also a spin-triplet state $|\uparrow,\uparrow\rangle$, $(|\uparrow,\downarrow\rangle + |\downarrow,\uparrow\rangle)/\sqrt{2}$ and $|\downarrow,\downarrow\rangle$. The singlet state is lower in energy than the triplet state due to the existence of inter-dot coupling, so that the ground state in the (1,1) region behaves as a singlet state with a minimum at $\phi = \pi$. These correspondences allow for the determination of the chain's occupancy number and parity by examining the ground-state EPR of the chains.

In addition, when the occupancy number on QD$_1$ is varied, two types of 0-$\pi$ transitions in the EPRs are expected, One is parity-flip 0-$\pi$ transition between (0,0) and (1,0), or between (0,1) and (1,1). The other is parity-preserving 0-$\pi$ transition between (0,0) and (1,1), or between (0,1) and (1,0).

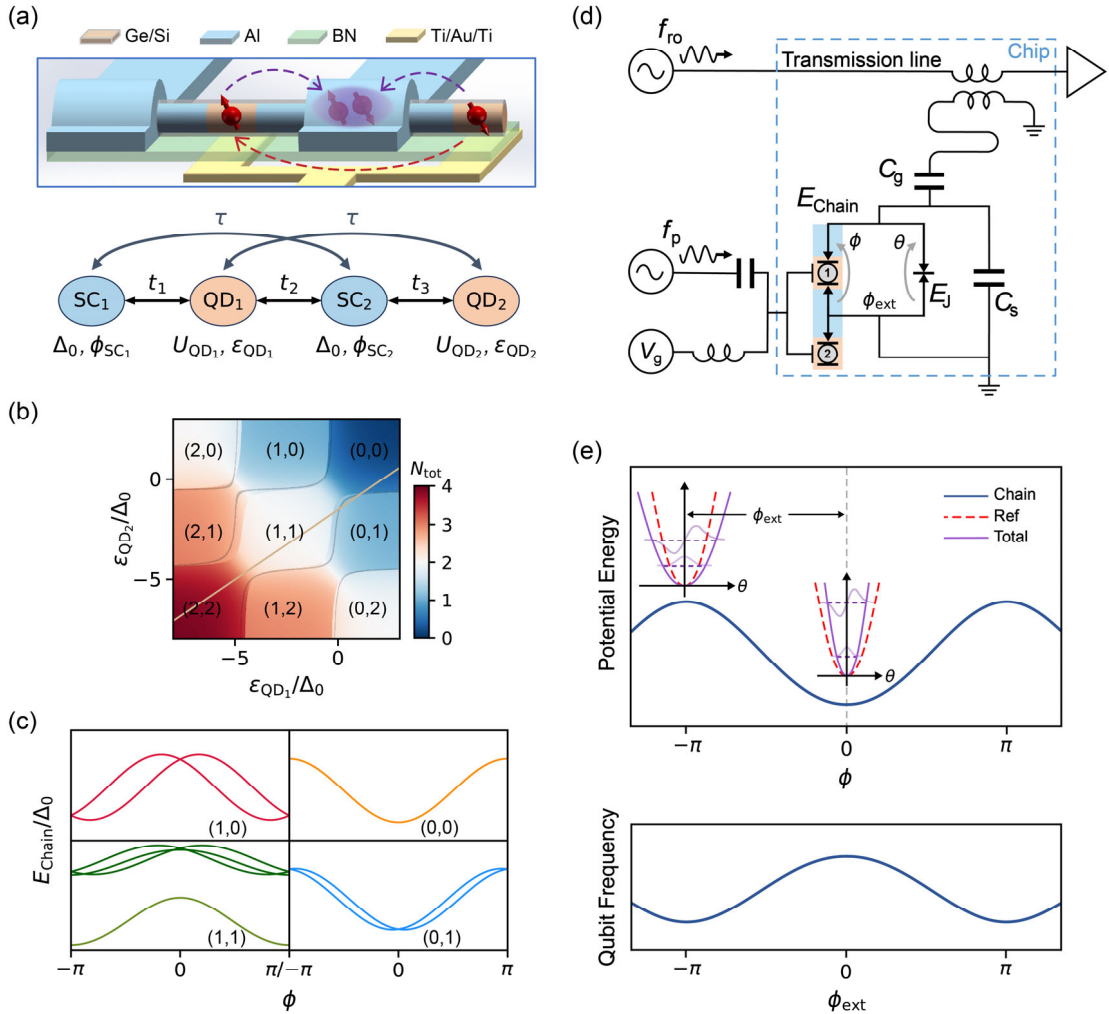

FIG. 1. Device and model. (a) Upper panel: Schematic illustration of a four-site SC-QD chain. The purple arrows indicate the CAR process, and the red arrow indicates the ECT process. Lower panel: The four-site model used in the analysis. The SC sites (with superconducting gap $\Delta_0$ and phase $\phi_{SC_i}$) and the QD sites (with Coulomb energy $U_{QD_i}$ and chemical potential $\varepsilon_{QD_i}$) are coupled by the nearest-neighbor hopping $t_j$ and the next-nearest-neighbor hopping $\tau$ ($i = 1, 2, j = 1, 2, 3$). For simplicity, we keep the two $\tau$ equal, and set $t_1 = t_2 = 0.5\Delta_0$, $t_3 = 0.8\Delta_0$, $U_{QD_1} = U_{QD_2} = 5\Delta_0$ and $\phi_{SC_2} = 0$ throughout the calculations. (b) The calculated charge stability diagram in the CAR-dominated regime ($\tau = 0.05\Delta_0$). The color scale represents the total occupation number ($N_{tot}$) of the chain at its ground states when the phase difference $\phi = \phi_{SC_1} - \phi_{SC_2} = 0$. The labels ($n_{QD_1}, n_{QD_2}$) denote the occupation numbers on each QD. The gray line denotes the parity boundary at $\phi = \pi$. (c) The expected EPRs of the SC-QD chain at four different charging states. (d) Measurement circuit diagram. An end of the chain is shunted to an Al/Al$_2$O$_3$/Al reference Josephson junction to form a transmon qubit. When an external magnetic flux $\Phi_{ext}$ is applied to the SQUID loop of the transmon, the system satisfies $\phi_{ext} = \phi - \theta$, where $\phi_{ext} = (2e/\hbar)\Phi_{ext}$ and $\theta$ is the phase difference across the Al/Al$_2$O$_3$/Al junction. (e) Schematic diagrams depicting the measurement principle. Upper panel: The phase dependent potential energy of the chain. Insets: The total potential energy of the transmon (purple solid lines) obtained by summing the potential energy of the chain (blue solid line) and the reference junction (red dashed lines). The total potential energy of the transmon is influenced by the chain's potential energy, leading to different qubit level spacings at different $\phi$. Lower panel: The resulted tansmon frequency as a function of $\phi_{ext}$, based on which the EPR and parity of the chain can be determined.

## Measurement circuit and principle —

To detect the EPR of the chain, we created a transmon qubit by incorporating an end of the chain with an Al/Al$_2$O$_3$/Al reference Josephson junction, as shown in Fig. 1(d). The superconducting quantum interference device (SQUID) loop of the transmon was shunted by a capacitor $C_s$ of ~100 fF, which provides a charging energy $E_C$ of ~200 MHz. And the SQUID loop was further coupled to a coplanar waveguide resonator through a capacitor $C_g$ of ~8 fF. The Josephson energy of the reference junction, $E_J$, was set to ~50 GHz, satisfying $E_J \gg E_C$, the criteria for a transmon qubit [38]. This designed $E_J$ was also considerably higher than that of the SC$_1$-QD$_1$-SC$_2$ junction on the chain, enabled the manipulation of the phase difference $\phi$ across the junction on the chain by adjusting the external magnetic flux $\Phi_{\text{ext}}$, thus $\phi_{\text{ext}}$. In addition, the electrochemical potentials in both QDs could be controlled by a shared electrostatic gate, which was fabricated using a 3/10/3 nm Ti/Au/Ti film stacked with a ~20 nm thick h-BN insulating layer. Further details about device fabrication and calibration can be found in the Supplemental Material.

In such a device, the total potential energy of the transmon, which is dominantly determined by the reference junction and the shunt capacitance, is modulated by the potential energy $E_{\text{Chain}}$ of the SC-QD chain, as shown in Fig. 1(e). In the weak $E_{\text{Chain}}$ case, the relationship between the transition energy of the transmon $\Delta E_{10}$ and the potential energy $E_{\text{Chain}}$ of the chain can be written as:

$$\Delta E_{10} = (\hbar\omega_{\text{p}} - E_{\text{C}}) + \frac{1}{2\lambda^2} \frac{\partial^2 E_{\text{Chain}}(\phi)}{\partial \phi^2}|_{\phi=\phi_{\text{ext}}} \quad (1)$$

where $\hbar\omega_{\text{p}} = \sqrt{8E_{\text{J}}E_{\text{C}}}$ and $\lambda = (E_{\text{J}}/8E_{\text{C}})^{1/4}$ (see Supplemental Material for detailed derivation). Since the first term in Eq. (1) remains constant, the flux dependence of $\Delta E_{10}$ is caused by the second derivative of $E_{\text{Chain}}$ at a given magnetic flux. This indicates that if $E_{\text{Chain}}$ exhibits a trigonometric dependence on the external flux, $\Delta E_{01}$ will behave similarly to the EPR of $E_{\text{Chain}}$, but with an opposite sign. Therefore, we can directly obtain the EPR of the SC-QD chain by measuring the energy spectrum of the transmon.

Leveraging its ability to access the chain's EPR, the transmon qubit serves as a parity-sensitive probe with several distinct advantages. Operating in the dispersive regime, it enables non-invasive measurements without quasiparticle injection, in contrast to tunneling-based methods. Moreover, the SC-QD chain is directly connected to the transmon's reference junction, providing stronger coupling than in capacitively coupled schemes. This leads to higher energy resolution and enables finer discrimination between different fermion parity states.

## Parity-flip 0-π transition—

We began by investigating the behavior of the device as the electrochemical potentials of the two QDs were tuned along the red line in the charge stability diagram shown in Fig. 2(a). Initially, the device was set to the (1,1) region, as indicated by the green triangle. A two-tone measurement was performed to measure the flux-dependent spectrum of the transmon. As shown in Fig. 2(b), the spectrum exhibits a branch with a maximum at $\phi_{\text{ext}} = \pi$. According to Eq. (1), this ground state of the chain corresponds to a $\pi$-phase-shifted singlet state. Through numerical simulations, we can obtain the EPRs of the chain at the position marked by the green triangle in Fig. 2(a). The results are shown in Fig. 2(c). It further supports the picture that the (1,1) state forms a singlet ground state due to the hybridization between the two QDs, with the minimum of EPR (thus the maximum in the transmon spectrum due to the sign change) shifted to the $\pi$ position because of the odd occupation in QD$_1$. We note that the parallel replicas in Fig. 2(b) were caused by multi-photon ac Stark effect [39].

Secondly, we set the device to the (0,1) charge configuration as indicated by the blue triangle in Fig. 2(a), and measured the spectrum of the transmon there. As shown in Fig. 2(f), the result reveals two horizontally shifted branches, each peaks near $\phi_{\text{ext}} = 0$. This observation agrees with the theoretical prediction of EPR of the chain in the (0,1) region (Fig. 2(g)). In this configuration, the ground state is characterized as a 0-phase-shifted state because the occupancy number of QD$_1$ is even. Due to the next-nearest-neighbor coupling and spin-orbit interaction, the potential energy of the chain is non-degenerate at magnetic fluxes away from the Kramers degeneracy point [40], leading to two slightly split and horizontally shifted doublet states.

Thirdly, we set the device to the (1,1) and (0,1) boundary in the charge stability diagram as marked by the cyan triangle in Fig. 2(a), and measured the two-tone spectrum of the transmon. Both singlet (1,1) state and doublet (0,1) state were observed, as illustrated in Fig. 2(d). Theoretically, the EPRs consist of both a $\pi$-phase-shifted singlet state and a 0-phase-shifted doublet state (Fig. 2(e)). The observation of coexistence is allowed if quasiparticle poisoning occurs. The quasiparticles may come from the band above the gap or from the environment [31, 41].

We also monitored the 0-$\pi$ transition continuously as a function of $\phi_{\text{ext}}$ and $V_{\text{g}}$ along the red line in Fig. 2(a), by utilizing the shift in the resonator's resonance frequency, $\Delta f_{\text{r}} = f_{\text{r}} - f_{\text{r,Ref}}$, as an indicator of the change in the transmon's resonance frequency (where $f_{\text{r,Ref}} = 6.48$ GHz is the resonator frequency set by the reference junction). The result is shown in Fig. 2(h). It can be seen that, with the increase of $V_g$, the ground state of the chain gradually evolves from a π-phase-shifted singlet state to 0-phase-shifted doublet states, with the coexistence of both states at intermedium values of $V_g$. During this process, the singlet ground state (S) (the green line) in Fig. 2(c) gradually lifts up, and the doublet states (D) (the blue lines) in Fig. 2(c) gradually go down. They overlap in energy with each other in Fig. 2(e). And finally, the doublet states become the new ground states in Fig. 2(g). The spectral weight in Figs. 2(b), (d) and (f) evolves accordingly. A movie for the evolution can be found in the Supplemental Videos. Note that the horizontal noise-like patterns in Fig. 2(h) arise from occasional instabilities and discrete jumps in the gate voltage during the slow sweep. These are likely caused by charge noise in the gate circuit or changes of trapped charges in the gate dielectric [42, 43].

We simulated the evolution of the Andreev levels of the chain as a function of $V_g$ at $\phi = 0$. As illustrated in Fig. 2(i), when the chemical potentials of the two QDs are tuned from the (1,1) region to the (0,1) region, the ground state of the chain transforms from the S(1,1) state to the D(0,1) states. Due to parity protection, there is no coupling between the even-parity S(1,1) state and the odd-parity D(0,1) state during this process. Nevertheless, since the quasiparticle poisoning rate in this experiment is significantly faster than the measurement speed, the measured spectrum is subject to fluctuations between different occupation states near the parity boundary, when the energy separation between the two occupation states is comparable to or smaller than the thermal fluctuation energy. Thus, the spectra of both states were detected simultaneously in Fig. 2(d).

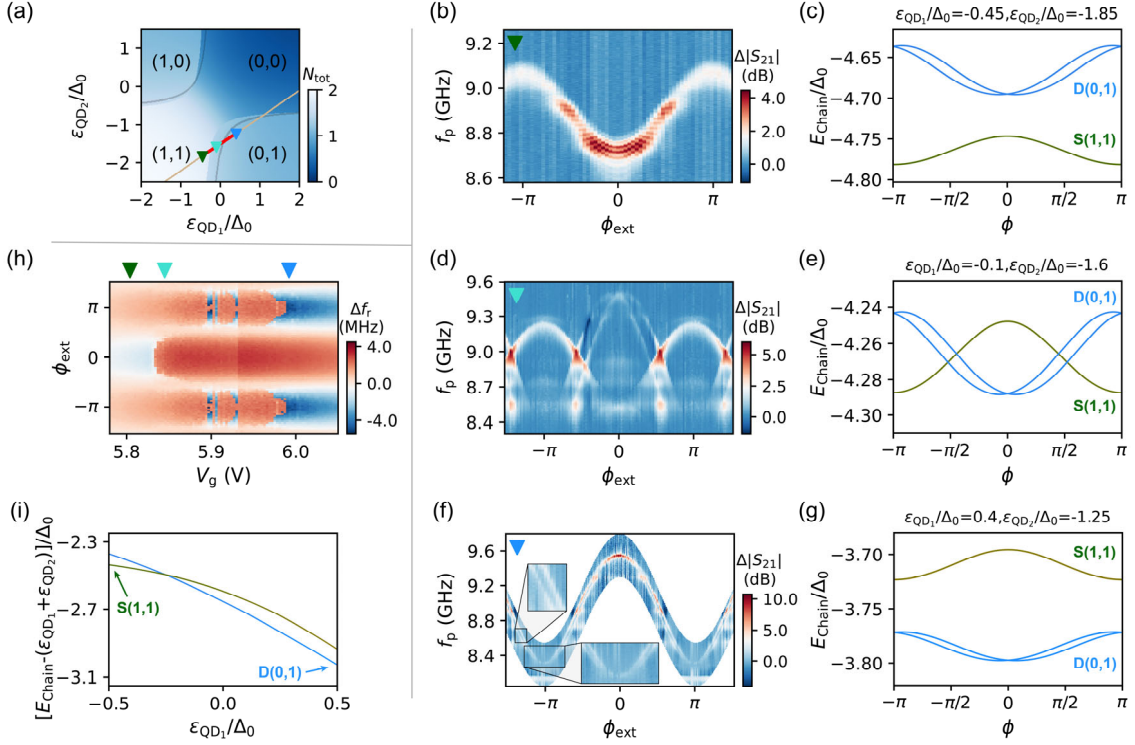

FIG. 2. Parity-flip 0-π transition. (a) A segment of the charge stability diagram in Fig. 1(b). The green, cyan and blue triangles denote the positions where the spectra of the transmon in (b), (d) and (f) were taken, respectively. The red line connecting these colored triangles represents the continuous tuning of $\varepsilon_{QD_1}$ and $\varepsilon_{QD_2}$ by $V_g$, along which the spectrum in (h) was taken. (b, d, f) The flux-dependent two-tone spectra of the transmon, taken at the $\varepsilon_{QD_1}$ and $\varepsilon_{QD_2}$ values indicated by the corresponding-colored triangles in (a). The insets in (f) show the details of two horizontally shifted doublet states. (c, e, g) The calculated EPRs of the Andreev levels of the chain, which are consistent with the experimental observations in (b), (d) and (f), respectively. (h) The spectrum of the resonator as a function of $\phi_{ext}$ and $V_g$, measured along the path indicated by the red line in (a), revealing the happening of a 0-π transition of the chain. The colored triangles indicate the $V_g$ values where the spectra in (b), (d) and (f) were taken. (i) The expected Andreev levels of the chain at $\phi = 0$, calculated along the path indicated by the red line in (a). The horizontal axis represents the changes in $\varepsilon_{QD_1}$, and the potential energy of the chain is shown after the background $\varepsilon_{QD_1} + \varepsilon_{QD_2}$ is subtracted. The green and blue lines indicate the potential energy of the even-parity S(1,1) and the odd-parity D(0,1) states, respectively.

## Parity-preserving 0-π transition—

We further studied the parity-preserving 0-π transition along the red line in Fig. 3(a), which crosses the zone between (2,2) and (1,1) regions where dominated CAR process happens in the chain. We initially set the gate voltage to 5.3 V, as marked by the orange star in Fig. 3(a), and measured the spectrum of the transmon. The result is shown in Fig. 3(b). The spectral line reaches its maximum frequency at $\phi_{ext} = 0$, indicating that the ground state of the chain was a singlet state without phase shifting. This measured spectrum is in agreement with the calculated S(2,2) ground state as shown in Fig. 3(c), confirming that the chain was in the (2,2) region. Again, the parallel replicas in Fig. 3(c) were caused by multi-photon ac Stark effect.

Subsequently, we set $V_g$ to 5.55 V marked by the green star in Fig. 3(a). In contrast to the orange star position, the transmon spectrum then exhibited a maximum at $\phi_{ext} = \pi$, as shown in Fig. 3(d). It indicates that this final ground state of the chain was a singlet state located in the (1,1) region, and that from the initial to the final ground states there involved a π-phase shift. This picture is supported by the calculated EPR of S(1,1) state shown in Fig. 3(e).

The evolution along the red line in Fig. 3(a) and the associated 0-π transition were further studied by measuring the shift of the resonator frequency $\Delta f_r$ as a function of $\phi_{ext}$ and $V_g$. The result is shown in Fig. 3(f). As $V_g$ increases, the line shape of the vertical linecuts in Fig. 3(f) gradually flattens out and then grows up with a π-phase shift (a movie can be found in the Supplemental Videos), implying that the ground state of the chain evolves continuously from a 0-phase-shifted state to a π-phase-shifted state without a change in parity.

In the next, we simulated the evolution of the states as a function of $V_g$ at $\phi = 0$. As shown in Fig. 3(g), when the two quantum dots are coupled, hybridization will occur between the states with same parity: i.e., the even-parity states S(2,2) and S(1,1) will hybridize at the degeneracy point, with an anti-crossing via CAR. Similarly, the odd-parity states D(1,2) and D(2,1) will hybridize and anti-cross due to ECT. Since the CAR process dominates in our chain, the hybrid states of D(1,2) and D(2,1) will always be the excited states in the hybridization region of the charge stability diagram. The ground state will undergo a continuous transition from the 0-phase-shifted S(2,2) state to the π-phase-shifted S(1,1) state without a change in the fermion parity, as observed in our experiment. On the contrary, if ECT was stronger than CAR, then the (1,2) and (1,2) regions in the charge stability diagram would connect with each other. For that case, when the chain was tuned from S(2,2) to S(1,1) by varying $V_g$, the ground state would undergo two parity changes, which obviously differs from our experimental observation [16].

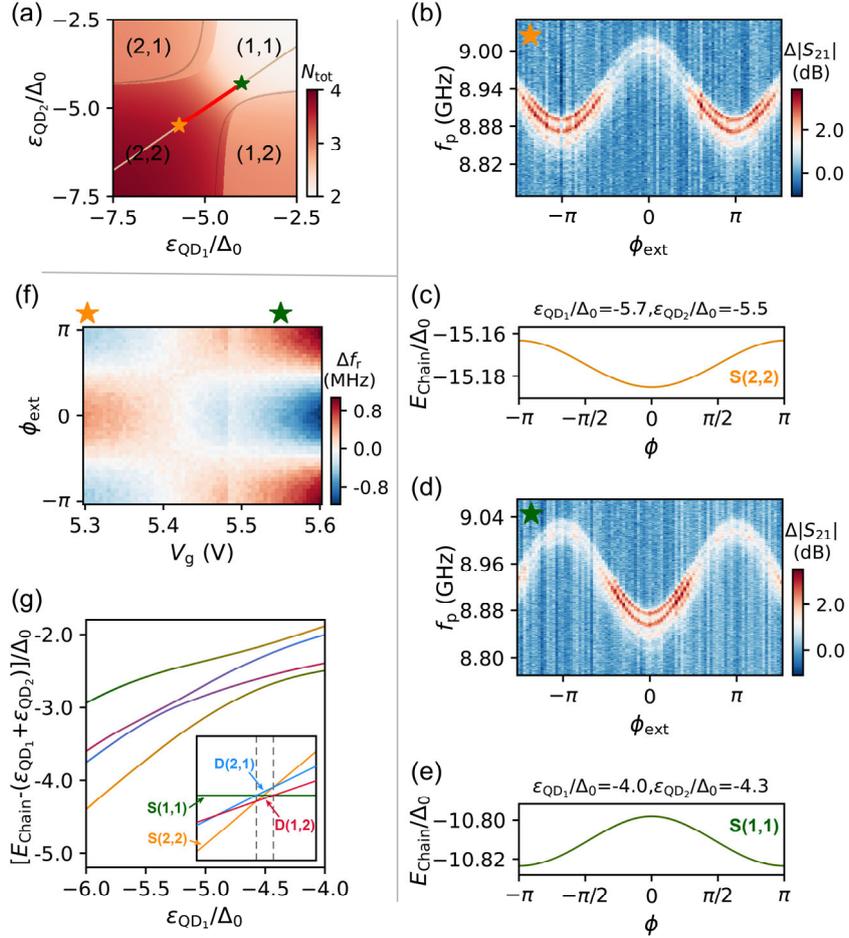

FIG. 3. Parity-preserving 0-π transition. (a) A segment of the charge stability diagram in Fig. 1(b). The orange and green stars denote the values of $\varepsilon_{QD_1}$ and $\varepsilon_{QD_2}$ where the spectra of the transmon in (b) and (d) were taken, respectively. The red line connecting the two stars represents the continuous tuning of $\varepsilon_{QD_1}$ and $\varepsilon_{QD_2}$ by $V_g$, along which the spectrum in (f) was measured. (b, d) The flux-dependent two-tone spectra of the transmon taken at the values of $\varepsilon_{QD_1}$ and $\varepsilon_{QD_2}$ indicated by corresponding-colored stars in (a). (c, e) The calculated EPRs of the Andreev levels of the chain, which are consistent with the experimental observations in (b) and (d), respectively. (f) The spectrum of the resonator as a function of $\phi_{\text{ext}}$ and $V_g$, measured along the path indicated by the red line in (a), revealing the happening of a 0-π transition of the chain. The colored stars indicate the $V_g$ values where the spectra in (b) and (d) were taken. (g) The expected Andreev levels of the chain at $\phi = 0$, calculated along the path indicated by the red line in (a). The horizontal axis represents the changes in $\varepsilon_{\text{QD}_1}$, and the potential energy of the chain is shown after the background $\varepsilon_{\text{QD}_1} + \varepsilon_{\text{QD}_2}$ is subtracted. The inset in (g) shows the Andreev levels in the absence of inter-sites coupling. The green, yellow, blue and red curves indicate the potential energy of the S(1,1), S(2,2), D(2,1) and D(1,2) states, respectively.

## Adjusting CAR and ECT strengths by electrostatic gating—

For the emergence of Majorana zero modes in a long Kitaev chain, it is crucial to own the capability to adjust the strengths of the CAR and ECT processes in the chain, such that the chemical potential at each QD is less than the hopping amplitudes, and that the hopping amplitudes of CAR and ECT between neighboring QDs follow the required sign order [12]. In the following, we demonstrate to use the electrostatic gate to adjust the strengths of CAR and ECT in the chain.

We examined the evolution of the states of the chain along the red line in the charge stability diagram shown in Fig. 4(a). At the lower-left beginning of the red line, the measured transmon spectrum of the D(0,1) state exhibited two horizontally shifted branches due to the strong spin-orbit interaction in Ge/Si nanowire, as shown in Fig. 4(b). The shifting between the two branches with opposite spin-polarization does not break the time-reversal symmetry.

At a fixed external magnetic flux away from the Kramers degeneracy points, i.e., at $\phi_{\text{ext}} = -0.46\pi$ as indicated by the black dashed line in Fig. 4(b), we measured the splitting of the D(0,1) states as a function of $V_g$. The results are shown in Fig. 4(c). By fitting the measured line shapes according to Eq. (1), with the $E_{\text{Chain}}$ term in Eq. (1) obtained by numerically solving the Hamiltonian of the chain, we obtained the blue dashed lines in the Fig. 4(c), together with the evolutions of the CAR and ECT strengths as a function of $V_g$, as shown in Fig. 4(d) (see Supplemental Material for derivation).

Here, both the CAR and ECT processes include the spin-parallel effective coupling ($\Gamma_{\uparrow\uparrow}^{CAR/ECT}$) and the spin-antiparallel effective coupling ($\Gamma_{\downarrow\uparrow}^{CAR/ECT}$).

In a spin-conserving system, CAR can only occur in a spin-antiparallel configuration, and ECT can only occur in a spin-parallel configuration. It is the one-dimensional Rashba spin-orbit interaction in the Ge/Si nanowire that enables the occurrence of spin-parallel CAR and spin-antiparallel ECT [12, 13, 44]. Notably, a weak magnetic field can produce a sufficiently strong Zeeman field to eliminate the antiparallel coupling term without disrupting superconductivity [45], thereby allowing the spin-polarized dot orbitals to serve as spinless fermions, which is required for the further construction of Kitaev chains. The curves of $\Gamma_{\uparrow\uparrow}^{CAR}$ and $\Gamma_{\uparrow\uparrow}^{ECT}$ in Fig. 4(d) indicate that one can vary the strength of CAR and ECT by tuning the gate voltage. In particular, it can be observed that the sign($\Gamma_{\uparrow\uparrow}^{CAR} \cdot \Gamma_{\uparrow\uparrow}^{ECT}$) remains positive within the range of the gate voltages explored, which is highly favorable for hosting Majorana zero modes in a long chain [12].

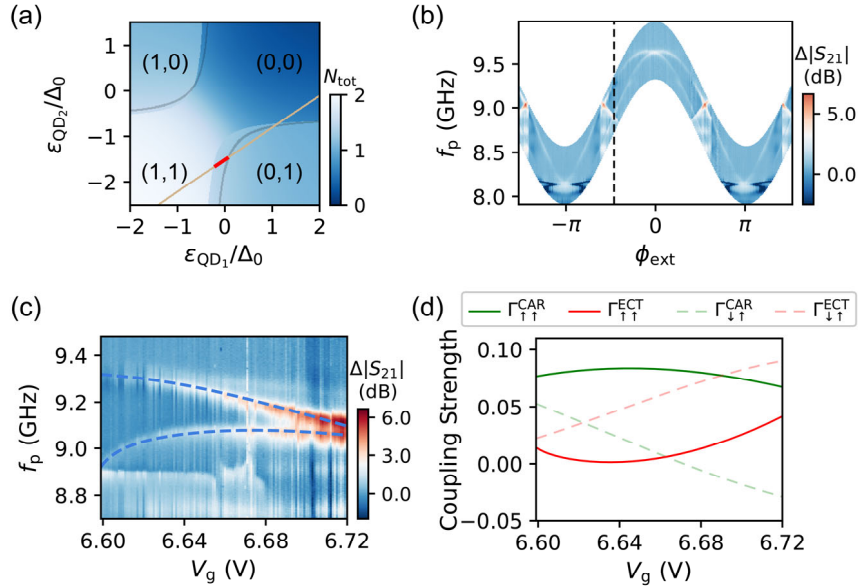

FIG. 4. The $V_g$ dependence of CAR and ECT. (a) A segment of the charge stability diagram in Fig. 1(b). The red line represents the continuous tuning of $\varepsilon_{QD_1}$ and $\varepsilon_{QD_2}$ by varying $V_g$, along which the spectrum in (c) was taken. (b) The flux-dependent two-tone spectrum of the transmon taken at the left end of the red line in (a). The black dashed line indicates the value of $\phi_{ext}$ where the spectrum in (c) was taken. (c) The $V_g$-dependent two-tone spectrum of the transmon taken along the path indicated by the red

line in (a), where the external magnetic flux was fixed at $\phi_{\text{ext}} = -0.46\pi$ as indicated by the black dashed line in (b). The blue dashed lines are the fitting curves to the transmon frequencies corresponding to the spin splitting of the D(0,1) state in the chain. (d) The $V_g$ dependence of the extracted strengths of CAR and ECT for the ↑↑ and ↓↑ spin configurations.

## Conclusion—

In summary, the EPR of the ground state in a Ge/Si nanowire-based four-site SC-QD chain was measured by using a transmon qubit. The singlet or doublet ground states, and thus the total parity of the chain, were determined. Combined with whether the ground state is 0-phase- or $\pi$-phase-shifted, the occupation number of charges in each QD can be further determined. We also demonstrated that the strengths of CAR and ECT in Ge/Si nanowire-based devices can be continuously tuned by electrostatic gating. This tunability, combined with the large spin-orbital coupling and the potential absence of nuclear spins in Ge/Si nanowires, makes this material a promising platform for constructing AKCs in the future.

In our experiment, the microsecond-scale coherence time of the transmon qubit ensures minimal spectral broadening, enabling the clear resolution of parity-dependent signatures in the EPR. These well-resolved spectral signatures suggest that future implementation of fast fermion parity readout via high-fidelity single-shot measurement of the transmon is feasible. As demonstrated in recent experiments on Andreev spin qubits [32-34, 46], an effective proposal is to distinguish different charge parity states—each associated with a distinct EPR—by mapping them to well-separated regions in the complex-plane histogram of the resonator response at a fixed readout frequency. We also note that when the transmon remains in its ground state, the readout circuit will not disturb the quantum state of the chain. The collapse of the parity superposition occurs only when a pump signal is applied to the transmon to probe the fused Majorana pair. These features collectively suggest that our approach could serve as a promising platform for real-time, projective measurement of fermion parity in artificial Kitaev chains.

## Acknowledgments—

This work was supported by the Innovation Program for Quantum Science and Technology through Grant No. 2021ZD0302600; by the National Natural Science Foundation of China (NSFC) through Grant Nos. 92365207, 92065203, 92365302, 11527806, 12074417, 11874406, 11774405 and E2J1141; by the Strategic Priority Research Program B of the Chinese Academy of Sciences through Grants Nos. XDB33010300, DB28000000, and XDB07010100; by the National Basic Research Program of China through MOST Grant Nos. 2016YFA0300601, 2017YFA0304700 and 2015CB921402; by Beijing Natural Science Foundation through Grant No. JQ23022; by Beijing Nova Program through Grant No. Z211100002121144; and by Synergetic Extreme Condition User Facility (SECUF).